\documentclass[12pt, prd, aps, preprintnumbers, nofootinbib, superscriptaddress]{revtex4}

\usepackage{epsfig}
\usepackage{amsmath}
\usepackage{amssymb}

\newcommand{\be}{\begin{equation}}
\newcommand{\ee}{\end{equation}}
\newcommand{\bee}{\begin{equation*}}
\newcommand{\eee}{\end{equation*}}
\newcommand{\bea}{\begin{eqnarray}}
\newcommand{\eea}{\end{eqnarray}}
\newcommand{\bean}{\begin{eqnarray*}}
\newcommand{\eean}{\end{eqnarray*}}

\newcommand{\nn}{\nonumber}

\newcommand{\tr}{\textrm{tr}}

\begin{document}

\preprint{UAB-FT-671}
\preprint{HD-THEP-09-20}

\title{Cold electroweak baryogenesis with Standard Model CP violation}

\author{A.~Tranberg}
\email{antranbe@sun3.oulu.fi}
\affiliation{Department of Physical Sciences, University of Oulu, P.O. Box 3000, FI-90014 Oulu, Finland}

\author{A.~Hernandez}
\email{a.hernandez@thphys.uni-heidelberg.de}
\affiliation{Institut f\"ur theoretische Physik,
Philisophenweg 16, 69120 Heidelberg, Germany }

\author{T.~Konstandin}
\email{konstand@ifae.es}
\affiliation{Institut de F\'isica d'Altes Energies,
Universitat Aut\`onoma de Barcelona, 08193 Bellaterra, Spain }

\author{M.~G.~Schmidt}
\email{m.g.schmidt@thphys.uni-heidelberg.de}
\affiliation{Institut f\"ur theoretische Physik,
Philisophenweg 16, 69120 Heidelberg, Germany }

\date{\today}

\begin{abstract}
We study a mechanism that generates the baryon asymmetry of the
Universe during a tachyonic electroweak phase transition. We utilize
as sole source of CP violation an operator that was recently obtained
from the Standard Model by integrating out the quarks.
\end{abstract}

\maketitle

%%%%%%%%%%%%%%%%%%%%%%%%%%%%%%%%%%%%%%%%%%%%%%%%%%%%%%%%%%%
\section{Introduction}
%%%%%%%%%%%%%%%%%%%%%%%%%%%%%%%%%%%%%%%%%%%%%%%%%%%%%%%%%%%%

The requirements for the dynamical creation of the baryon asymmetry in
the Universe are stated by the Sakharov conditions: Violation of
baryon number conservation, violation of charge conjugation (C) and
charge-parity conjugation (CP) symmetry, and departure from
equilibrium~\cite{Sakharov:1967dj}. While there exists a plethora of
possible explanations for the baryon asymmetry, for a review see
e.g.~\cite{Dine:03}, all of them have to introduce physics beyond the
Standard Model (SM) in order to provide all these ingredients. 
Besides, most of them operate at energy ranges that are not
testable by experiments in the near future, and hence become difficult
to falsify. In principle, the SM contains all three necessary
ingredients mentioned before. Baryon number violation is supplied by
the weak anomaly \cite{tHooft:1976up} and weak interactions violate C
maximally and violate CP through the Kobayashi-Maskawa mechanism
\cite{CKMmatrix:1973}. Finally, departure from equilibrium could occur
due to expanding Higgs bubbles during a first-order electroweak phase
transition, and the corresponding mechanism is called electroweak
baryogenesis~\cite{Kuzmin:1985mm}. However, the experimental lower
bound on the Higgs mass implies that there did not occur a strongly
first-order phase transition as required but a crossover in the
SM~\cite{Kajantie:97}. Even if the phase transition would be
first-order, the CP violation from the CKM matrix is not strong enough
in the Higgs bubble walls to generate a substantial baryon asymmetry.
While this rules out electroweak baryogenesis in the SM, there are
extensions of the SM which overcome these shortcomings while still
remaining close to the SM. This includes e.g.~the MSSM with CP
violation in the chargino~\cite{Carena:2002ss, Konstandin:2005cd} or
neutralino sector~\cite{Cirigliano:2006dg} or singlet extensions of
the MSSM~\cite{Huber:2000mg, Huber:2006wf}.

Another scenario is given by the so-called cold electroweak
baryogenesis where the electroweak phase transition is tachyonic and
initiated at the end of (inverted) low-scale hybrid inflation. In this
case, not the temperature dependence of the free energy but the
inflation field turns the effective squared Higgs mass parameter
negative~\cite{Krauss:1999ng, GarciaBellido:1999sv, Copeland:2001qw,
GarciaBellido:2003wd, Smit:2002sn, Tranberg:2003gi}. The electroweak
phase transition occurs basically at zero temperature, and because of
the spinodal instability induced on the Higgs field, its low momentum
modes grow exponentially~\cite{Felder:2000hj, GarciaBellido:2002aj,
Smit:2002sn}. This allows for a classical treatment of the dynamics.

The generation of the baryon asymmetry in this scenario has been
simulated on the lattice~\cite{Tranberg:2003gi}, where the source of
CP violation was assumed to be of the form
\be
\label{eq:CPold}
\frac{\kappa_{CP}}{M^2}\phi^{\dagger}\phi \, 
\epsilon^{\mu\nu\lambda\sigma} \,
\tr \, \left(W_{\mu\nu} W_{\lambda\sigma}\right),
\ee
where $W$ denotes the $SU(2)_L$ field strength. A term of this form
could originate in an effective action from a more fundamental theory
at higher energies or from integrating out heavy fermions. In the
present work we report on the simulation of cold electroweak
baryogenesis using a CP-violating operator that was recently obtained
by integrating out the quarks of the Standard
Model~\cite{Hernandez:2008db} and reads
\bea
\label{eq:CPterm0}
\frac{\kappa_{CP}}{M^2}
\epsilon^{\mu\nu\lambda\sigma}
\biggl(Z_{\mu}W^{+}_{\nu\lambda}W^{-}_{\alpha}
\left(W^{+}_{\sigma}W^{-}_{\alpha}+W^{+}_{\alpha}W^{-}_{\sigma}\right)
+ \, c.c.\biggr).
\eea
Our main concern is if an operator of this form can successfully bias
the Chern-Simons number and through the anomaly, baryon number, 
during tachyonic preheating and hence explain
the observed baryon asymmetry via cold electroweak baryogenesis.

It has often been argued in the literature that CP violation in the SM
is too small to be able to generate the baryon asymmetry and we review
this argument and the results of ref.~\cite{Hernandez:2008db} in the
next section. In Sec.~\ref{sec_num} we present numerical results of
the lattice simulation of cold electroweak baryogenesis.  Finally, in
Sec.~\ref{sec_con} we conclude.

%%%%%%%%%%%%%%%%%%%%%%%%%%%%%%%%%%%%%%%%%%%%%%%%%%%%%%%
\section{CP violation in the Standard Model \label{sec_cpv}}
%%%%%%%%%%%%%%%%%%%%%%%%%%%%%%%%%%%%%%%%%%%%%%%%%%

It is often stated in the literature that the CP violation present in
the SM is insufficient to explain the observed baryon asymmetry. These
claims rest usually on the so-called Jarlskog
determinant~\cite{Jarlskog} and we review this argument in the
following. The basic observation is that physical observables cannot
depend on the flavor basis chosen for the quarks; in particular
transformations of the right-handed quarks leave the Lagrangian
invariant since the weak interactions are chiral. Besides, the quark
fields can be redefined absorbing one complex phase. The last fact
implies that all CP-odd observables in the SM have to be proportional
to
\be\label{eq:valJ}
J=s_1^2s_2s_3c_1c_2c_3 \sin(\delta)=(3.0\pm0.3)\times10^{-5},
\ee
with the Jarlskog invariant $J$ given in terms of the
Kobayashi-Maskawa parametrization of the CKM matrix $V$ with a
CP-violating phase $\delta$ as defined in refs.~\cite{Jarlskog,
Yao:2006px}. In addition, if two up- or down-type quark masses were
degenerate, there would be no CP violation in the Standard Model since
flavor basis transformation can in this case be used to remove the
complex phase of the CKM matrix altogether from the Lagrangian.

If one further assumes that the observable under consideration is
polynomial in the quark masses, the simplest dimensionless expression
that fulfills these constraints is found to be the Jarlskog
determinant that has the form
\bea
\label{eq_def_Jinv}
\Delta_{CP} &=& v^{-12} \textrm{Im  Det} \left[ m_u m_u^\dagger, 
m_d m_d^\dagger \right] \nn\\
&=& J \, v^{-12} \, \prod_{i<j} (\tilde m_{u,i} - \tilde m_{u,j}^2)
\prod_{i<j} (\tilde m_{d,i}^2 - \tilde m_{d,j}^2)
\simeq 10^{-19},
\eea
where $v$ is the Higgs vacuum expectation value and $\tilde m^2_{u/d}$ denote the diagonalized mass matrices
according to
\be
m_d m_d^\dagger = D \tilde m^2_d D^\dagger, \quad
m_u m_u^\dagger = U \tilde m^2_u U^\dagger.
\ee
The identity in eq.~(\ref{eq_def_Jinv}) results then from the
following relation of the CKM matrix (summation over indices is only
performed as explicitly shown)
\be
\label{eq_ckm_sum}
\textrm{Im} \left[ V_{ab} V^\dagger_{bc} V_{cd} V^\dagger_{da} 
\right]
= J \sum_{e,f} \epsilon_{ace} \epsilon_{bdf}, \quad
V= U^\dagger D.
\ee

According to this argument CP violation in the SM seems to be too
small to explain the observed baryon asymmetry that is of order
$\eta\sim 10^{-10}$ and several proposals in the literature aim at
avoiding this bound. For example, it has been argued that at
temperatures of the electroweak scale the CP violation might be only
suppressed by the temperature rather than by the Higgs vev $v$ as
given in~eq.~(\ref{eq_def_Jinv}), but this still is insufficient to be
significant in a baryogenesis
mechanism~\cite{Farrar:1993sp,Farrar:1993hn}. In the context of
electroweak baryogenesis coherent scattering at the bubble wall has
been suggested in the same work~\cite{Farrar:1993sp,Farrar:1993hn} but
finally dismissed~\cite{Gavela:1994}. Furthermore construction of
rephasing invariants containing derivatives of the Higgs field have
proved to be incapable of being relevant in electroweak
baryogenesis~\cite{Konstandin:2003dx}.

The approach in refs.~\cite{worldline, Hernandez:2008db} is based on
the effective action obtained by integrating out the quarks of the SM
in the gradient expansion (more precisely an expansion in Lorentz
indices, i.e. the covariant derivative expansion). The resulting effective action can potentially contain
factors of the form
\be
\frac{m_i^2 - m_j^2}{m_i^2 + m_j^2}, \quad
\log[m_i^2/m_j^2].
\ee
These expressions vanish for degenerate quarks but are unlike the
factors in the Jarlskog determinant not suppressed for small Yukawa
couplings. The appearance of terms of this form was already argued in
the context of cold electroweak baryogenesis in
ref.~\cite{Smit:2004kh}. Indeed when the quarks of the SM are
integrated out, one finds in the effective action in next-to-leading
order of the gradient expansion the operator~\cite{Hernandez:2008db}
\bea\label{eq:CPterm}
\frac{i}{8(4\pi)^2}\frac{N_c}{16} \frac{J\,\kappa_{CP}}{\tilde m_c^2}
\epsilon^{\mu\nu\lambda\sigma}\int d^4x\left(\frac{\upsilon}{\phi}\right)^2
\biggl(Z_{\mu}W^{+}_{\nu\lambda}W^{-}_{\alpha}
\left(W^{+}_{\sigma}W^{-}_{\alpha}+W^{+}_{\alpha}W^{-}_{\sigma}\right)
+ \, c.c.\biggr),
\eea
where $J$ is the Jarlskog invariant given in eq.~(\ref{eq:valJ}) and
\be\label{kappa}
\kappa_{CP} \approx 9.87.
\ee
As required, the operator is proportional to $J$ and would vanish for
degenerate quark masses. However, the latter information is hidden in
the numerical coefficient $\kappa_{CP}$ that is a function of the six
quark masses
\footnote{ There is a recent 
claim~\cite{GarciaRecio:2009zp} that an alternative method to the one
used in ref.~\cite{Hernandez:2008db} leads to no CP violation in the
imaginary part of the Euclidean effective action in next-to-leading
order. Moreover, using this method all operators considered by the
authors (all CP-odd and several CP-even operators) of the imaginary
part apparently vanish to this order in four dimensions. On the other
hand, the next-to-leading order result in two dimensions (as presented
in \cite{worldline} and confirmed with a different method in
\cite{Salcedo:2008bs}) does not vanish. A vanishing result in four
dimensions seems implausible. We also stress that the results in
\cite{Hernandez:2008db} are obtained on par with a large number of
consistency checks and computed with the help of a computer algebra
program.}
\footnote{The same work~\cite{GarciaRecio:2009zp} also reported on 
CP violation in the real part of the Euclidean effective action that
is however subdominant for cold electroweak baryogenesis due to parity
conservation.}. 
The function is symmetric under the exchange of two
families but finite in the limit of one family becoming massless. The
scale of the operator is hence given by the second heaviest family.

This result was obtained in unitary gauge and the action can be
rewritten in $SU(2)_L$ gauge invariant quantities. The
charged gauge fields can be rewritten as
\be\label{eq:Wtogauge}
W_{\mu\nu}^+ = \frac{\phi^\dagger W_{\mu\nu} \tilde \phi}{ \phi^\dagger \phi}, \quad
W_{\mu\nu}^- = \frac{{\tilde \phi}^\dagger W_{\mu\nu} \phi}{ \phi^\dagger \phi}, \quad
W_{\mu}^+ = \frac{\phi^\dagger {\cal D}_\mu \tilde \phi}{ \phi^\dagger \phi}, \quad
W_{\mu}^- = \frac{{\tilde \phi}^\dagger {\cal D}_{\mu} \phi}{ \phi^\dagger \phi}, \quad
\ee
and similarly for the uncharged quantity
\be\label{eq:Ztogauge}
Z_\mu = W_\mu^3 - B_\mu = \frac{\phi^\dagger {\cal D}_{\mu} \phi  
- {\tilde \phi}^\dagger {\cal D}_{\mu} \tilde\phi }
{ 2\phi^\dagger \phi}.
\ee

In the following we employ the operator (\ref{eq:CPterm}) in a
modified form in the lattice study of cold electroweak
baryogenesis. This is necessary, since the gradient expansion used to
obtain the result is only strictly valid in the case
\be
\label{eq_applicability}
Z_\mu, W^\pm_\mu,  \ll m, \quad Z_{\mu\nu}, W^\pm_{\mu\nu} \ll m^2.
\ee
As mentioned earlier, the operator turns
out to be finite for vanishing up and down quark
masses and one finds that the next-to-leading order operator is
suppressed by the charm mass as indicated in the notation of
eq. (\ref{eq:CPterm}). It is plausible that this is the appropriate
mass scale that has to be used in the inequalities
(\ref{eq_applicability}). Still, the vev of Higgs field vanishes at
certain points during the tachyonic phase transition, thus
invalidating the expansion. 

In order to avoid that the dynamics is dominated by those points, we
introduce a cutoff. The purpose of the cutoff is to suppress the CP
violation in the region where the operator~(\ref{eq:CPterm}) is
artificially large due to the break down of the gradient expansion.
For the mechanism of cold electroweak baryogenesis, it is not
essential that CP violation is active for Higgs configurations with
almost vanishing vev. Two configurations with CP conjugate initial
conditions will slowly drift apart, eventually leading to a difference
in Higgs winding number, even if CP violation is only active in the
regions of large Higgs vev.

Instead of an expansion in gradients, an expansion in inverse
gradients is feasable in the regions of small Higgs vev and higher
order operators should come with a suppression of order
\be
\label{eq:cutoff_est}
\Lambda^2 \approx \frac{(\partial_\mu \phi)^2}{v^2} 
\lesssim \frac{V_{pot}}{v^2} \approx \frac{m_H^2}{8}.
\ee
In practice, we insert $\Lambda$ whenever the Higgs field appears in
the denominator
\be
\label{eq:cutoff}
\frac{1}{\phi^\dagger \phi} \to \frac{1}{c(\phi^\dagger \phi + \Lambda^2)}, 
\qquad c\left(\frac{v^2}{2}+\Lambda^2\right)=\frac{v^2}{2}.
\ee
and $c$ is just a constant fixed to satisfy the last condition.  Even
after introduction of the cutoff, the gradient expansion could be
jeopardized by too large gauge fields. Means of justifying the
gradient expansion could be to analyze the dependence of our results
on the cutoff and the impact of the CP-violating operator {\em a
posteriori}.

%%%%%%%%%%%%%%%%%%%%%%%%%%%%%%%%%%%%%%%%%%%%%%%%%%
\section{Numerical analysis \label{sec_num}}
%%%%%%%%%%%%%%%%%%%%%%%%%%%%%%%%%%%%%%%%%%%%%%%%%%%%

Following the analysis of ref.~\cite{Ambjorn:1990pu,
vanderMeulen:2005sp, Tranberg:2003gi, Tranberg:2006ip,
Tranberg:2006dg}, the SU(2)-Higgs model is discretized on a space-time
lattice, and the classical equations of motion derived in a
straightforward way. The implementation of the CP-violating operator
hereby requires symmetrization in space and time, leading to implicit
equations of motion in time, again in a similar fashion as for the
operator (\ref{eq:CPold}) used in~\cite{Tranberg:2003gi}. The
equations are solved by iteration and convergence of this iteration
procedure imposes certain restrictions on the coefficient
$\kappa_{CP}$, the timestep $dt$ and the cutoff $\Lambda$, in order
for the CP-violating force to not be too large. Ultimately, we are
interested in the Higgs winding number and the Chern-Simons number
that read
\bea
N_w &=& \frac{1}{24 \pi^2} \epsilon_{ijk}\int dx^3
(\phi^\dagger\partial_i\phi)(\phi^\dagger\partial_j\phi)
(\phi^\dagger\partial_k\phi), \\
N_{\rm cs} &=& \frac{1}{32 \pi^2} \epsilon_{ijk}\int dx^3
\left( W_i^a W_{jk}^a - \frac13 \epsilon_{abc} W_i^a W_j^b W_k^c \right).
\eea
where we used the temporal gauge, $W_0^a = 0$, and $W^a_{\mu\nu}$
denotes the corresponding field strength.

The simulation generates random configurations on a $n_x^3=64^3$
lattice, with lattice spacing $am_H=0.35$. These are evolved using the
equations of motion to $m_H t=30$, with timestep $dt=0.0125$. Each
configuration is run twice, using $\kappa_{CP}=\pm 50$, and subtracting the
resulting $N_{\rm cs}$ and $N_{\rm w}$ values. This way, the ensemble
of initial conditions is CP symmetric and the asymmetry is strictly
zero at $\kappa_{CP}=0$. This drastically reduces statistical noise,
allowing a much clearer
signal~\cite{Tranberg:2006ip}. Correspondingly, only instances with
integer difference in the Chern-Simons number between $\pm \kappa_{CP}$
contribute.  Statistics of this sort of observable is binomial such
that the standard deviation is given by
\be
\sigma=\frac{\sqrt{\frac{N_{\rm jumps}}{2^2N_{\rm configs}}-\left(\frac{N_{\rm jumps}}{2N_{\rm configs}}\right)^2}}{\sqrt{N_{\rm configs}-1}}.
\ee
It will turn out that the asymmetry for all pairs of configurations is
non-negative and that the result is overall not consistent with zero.

\begin{figure}
\includegraphics[width=0.45\textwidth, clip ]{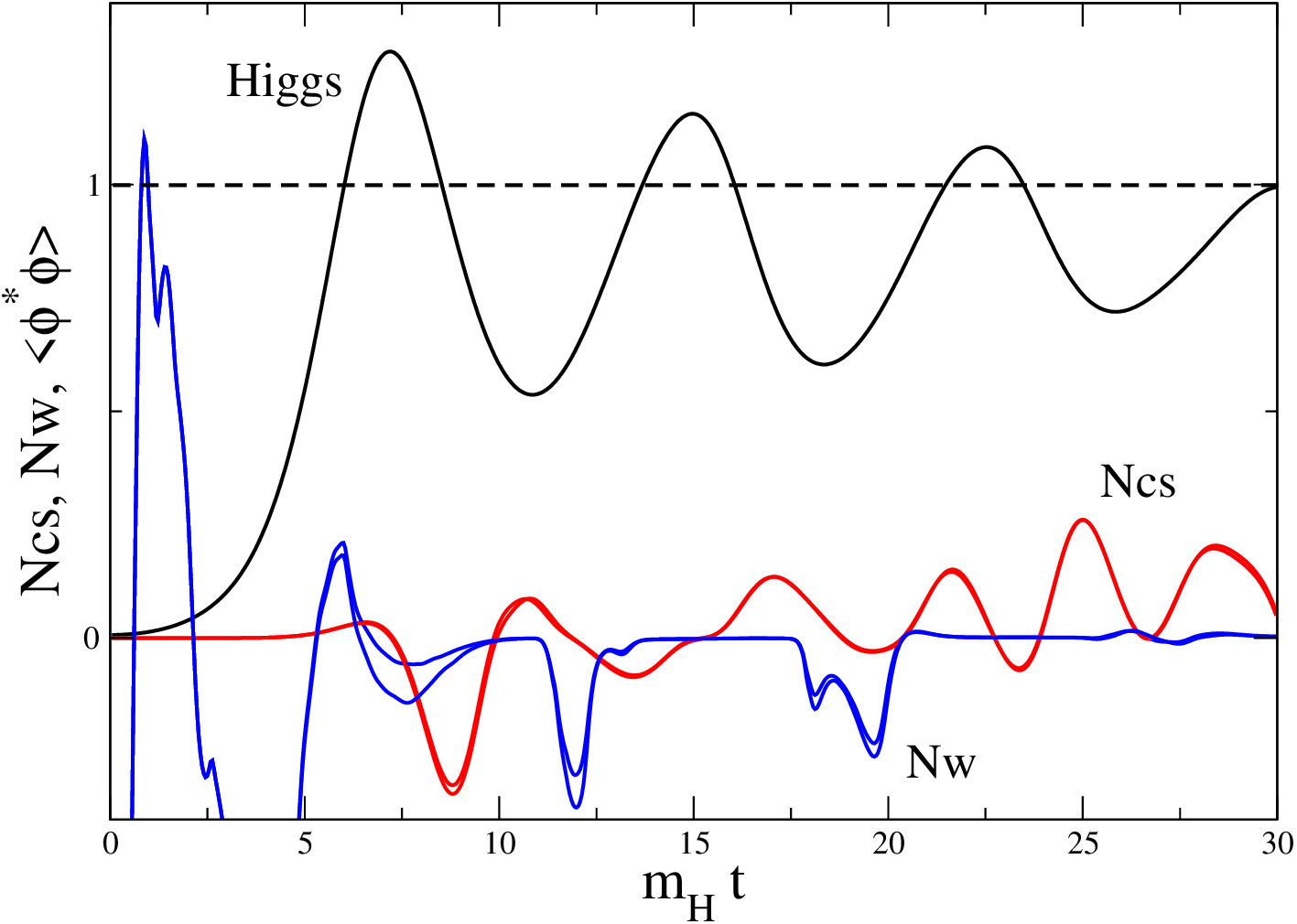}
\includegraphics[width=0.45\textwidth, clip ]{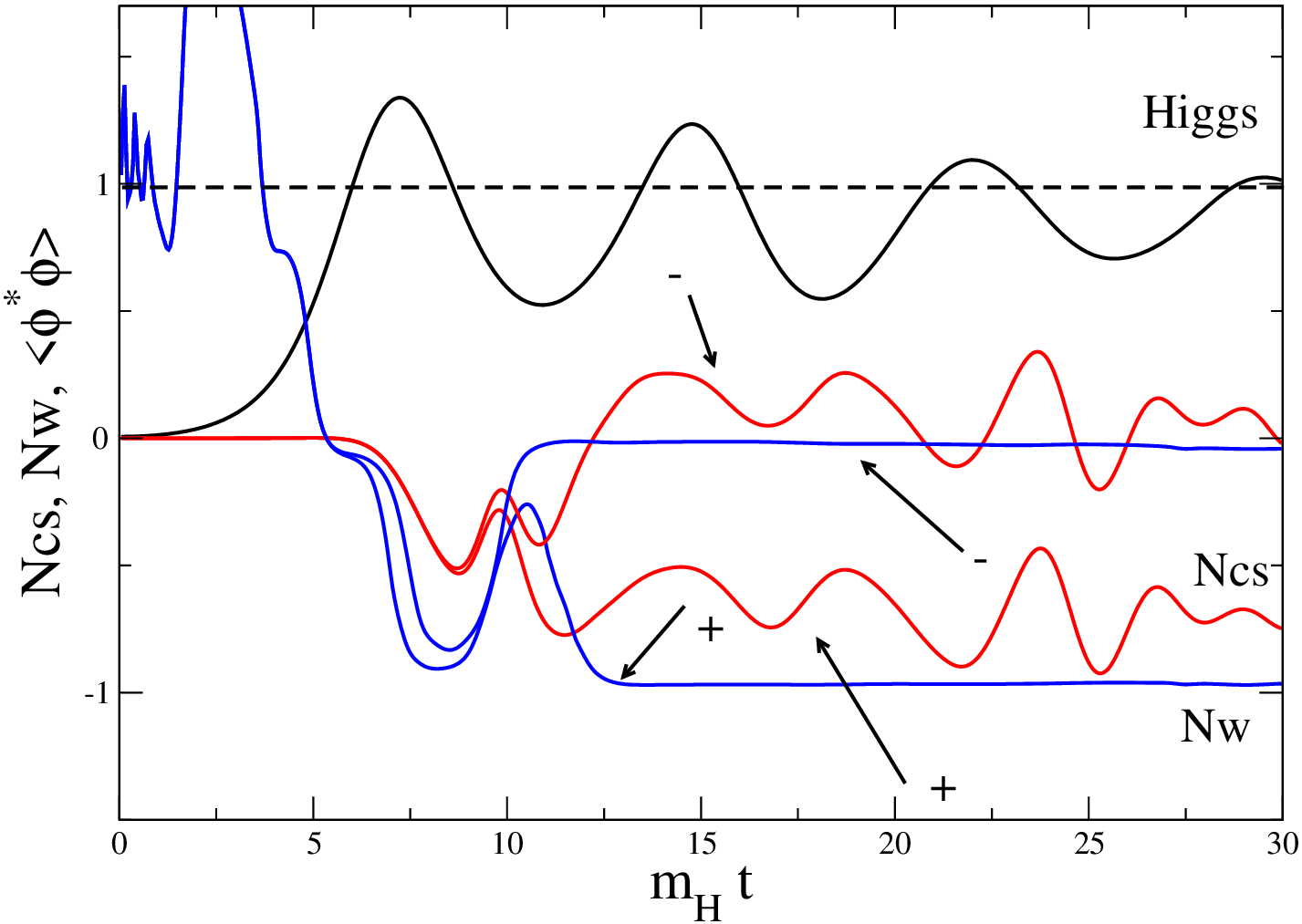}
\caption{\label{fig:jump}
Two example configuration: The right configuration results in a
difference in the Higgs winding number for CP conjugate initial
conditions, while the left does not. The Higgs field (straight curve)
is normalized to $v^2$. The red and blue lines denote $N_{cs}$ and
$N_w$ respectively.}
\end{figure}

As parameters we use $m_H=2m_W$ and a cut-off of $c\Lambda^2=50^2$ GeV$^2$.
The value of the cutoff corresponds to the estimate in 
(\ref{eq:cutoff_est}). In the simulation, we observe that the baryon
asymmetry scales roughly with the fourth power of the cutoff what
introduces a sizable uncertainty. A more sophisticated estimate of the
cutoff and a more extensive presentation of the data will be given
elsewhere~\cite{Tranberg:inprep}.

Figure \ref{fig:jump} shows two example configurations, one of which
contributes to the asymmetry. The Higgs field falls into the broken
phase and performs a damped oscillation. In the meantime, energy is
transferred into the gauge and Higgs fields, and their modes grow
under influence of the CP violation.

A first difference between the CP conjugate configurations occurs
typically around $t m_H \approx 6$. At this time already a substantial
amount of energy is transformed into the gauge fields and the Higgs
vev is non-vanishing what is essential for the CP-violating operator
(\ref{eq:CPterm0}) to be relevant. At the first minimum of the Higgs
oscillation, where many zeros of the Higgs field are
created~\cite{vanderMeulen:2005sp} winding number is potentially
generated. The presence of CP violation can hereby lead to a net
baryon number between configurations with CP conjugate initial
conditions. At late times winding number and Chern-Simons number agree
and in the case at hand the Chern-Simons number follows the previously
generated winding number. At the end, winding number approaches an
integer valued vacuum while Chern-Simons number, containing thermal noise, oscillates for a
longer time before it eventually settles into the same integer value~\cite{Tranberg:2003gi}.

Notice that this behavior is quite different to the mechanism based on
the operator (\ref{eq:CPold}). For a non-vanishing Higgs field this
operator can be interpreted as a chemical potential for the
Chern-Simons number. Accordingly, Chern-Simons number is already
generated during the first roll-off of the Higgs field and in the
first minimum of the Higgs field the winding accommodates to the
Chern-Simons number instead of vice-versa~\cite{Tranberg:2006ip}.

\begin{figure}
\includegraphics[width=0.7\textwidth, clip ]{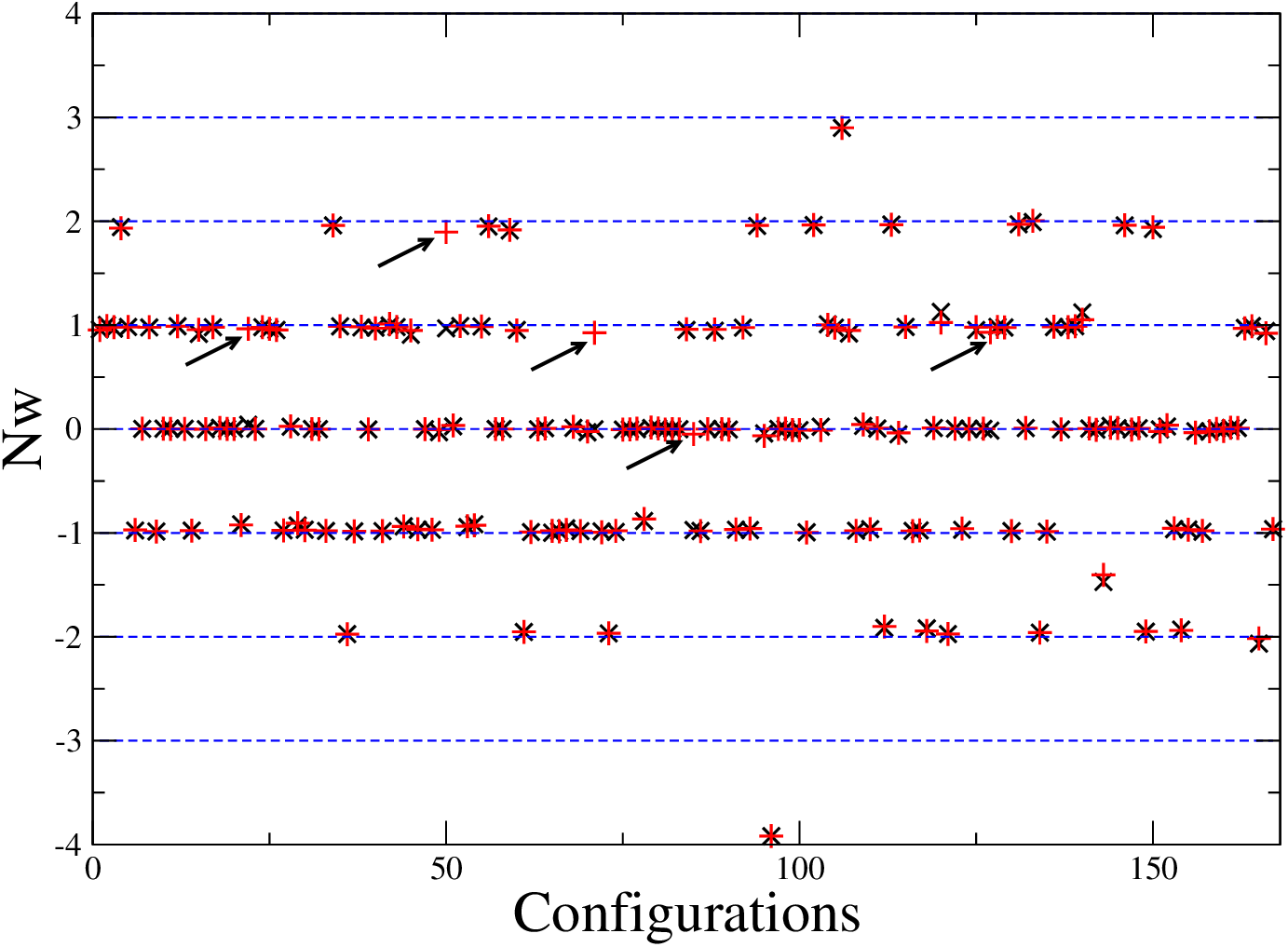}
\caption{\label{fig:allconf}
The final value of Higgs winding number for all $167$ pairs of
configurations. Black $x$-marks correspond to $+\kappa_{CP}$, red $+$-marks
to $-\kappa_{CP}$. In total $5$ pairs of configurations contribute to a net
baryon asymmetry. Deviations from integer values are rather small and
due to lattice discretization errors.}
\end{figure}

The final value of winding number is shown for an ensemble of $167$
configurations in fig.~\ref{fig:allconf}.  We note that all final
values are very close to integers, showing that the lattice
discretization errors are well under control. Only for one pair of
configurations the winding number did not settle yet in an integer
value at the end of the simulation. There are five configurations with
a net winding number, indicated by arrows. Hence the asymmetry is
\be
\delta N_{\rm w}=\frac{5}{334}\pm \frac{5}{334}
\sqrt{\left(1/5-1/167\right)\frac{167}{166}}=0.015\times(1\pm0.44).
\ee

\begin{figure}
\includegraphics[width=0.7\textwidth, clip ]{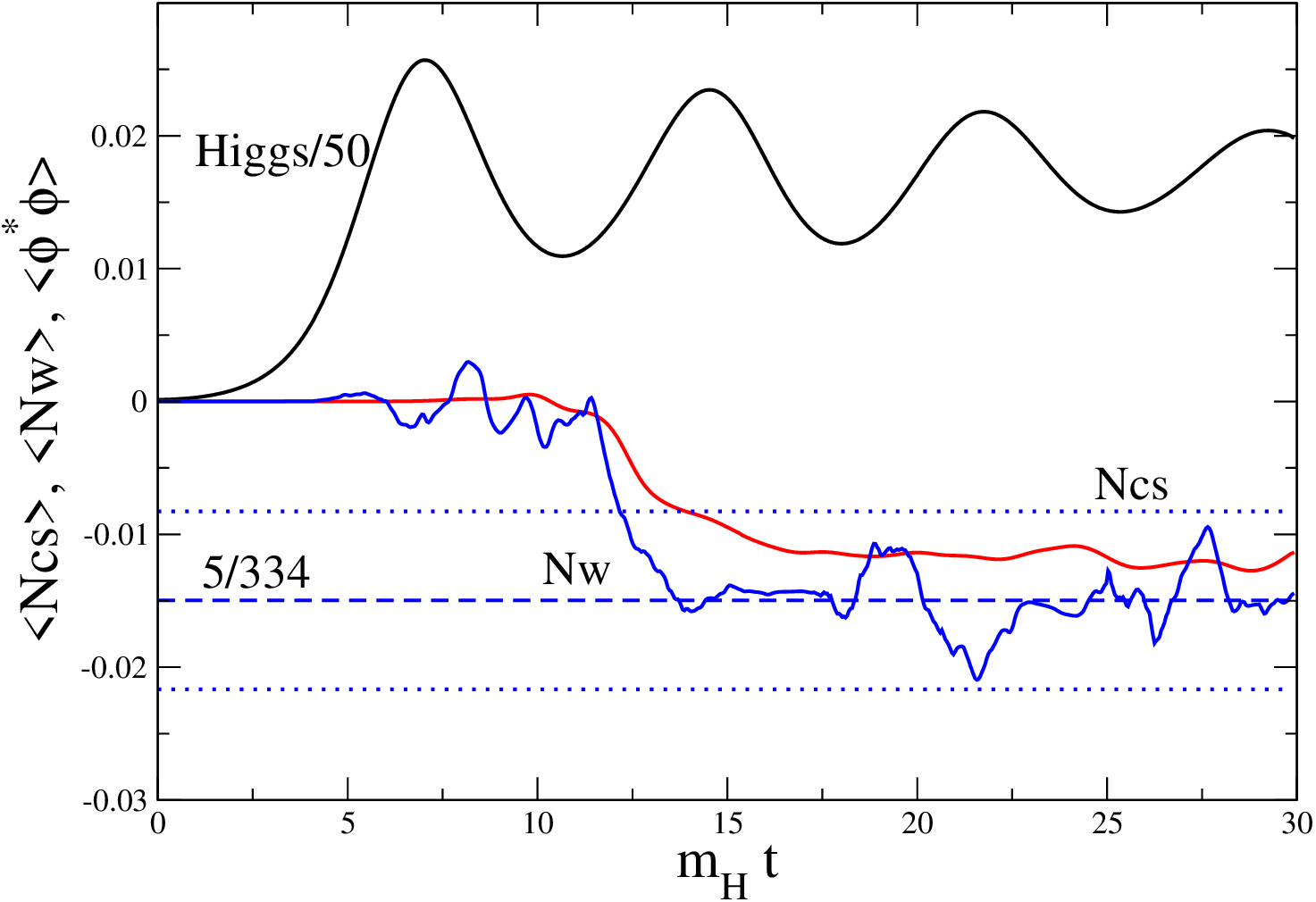}%figs/
\caption{ \label{fig:average}
The average Higgs field (scaled by a factor $1/50$), Chern-Simons number
and winding number as functions of time. The average value $5/334$ is
superposed (dashed line) as well as the estimated error band (dotted
lines).  }
\end{figure}

Figure \ref{fig:average} shows the average of Higgs field,
Chern-Simons number and Higgs winding numbers as functions of time. The
average value $5/334$ and the error estimate are indicated by blue
dashed and dotted lines, respectively. The same picture arises as for
the example configurations: Chern-Simons and winding numbers vanish
until the first Higgs minimum, where winding number is generated
potentially in a CP-violating way. The Chern-Simons number
accommodates to the winding number subsequently.

The baryon number density, reheating temperature and photon number
density are given by
\be
n_B=3n_{\rm cs},\quad V_0=\frac{\lambda v^4}{4}
=\frac{\pi^3 g^*}{30}T^4_{\rm reh},\quad n_\gamma
=\frac{1}{7}\frac{2\pi^2 g^*}{45}T^3_{\rm reh}.
\ee
For the Standard model $g^*=86.25$, and $v=246$ GeV and we choose $m_H=2m_W\simeq
160$ GeV, what yields for the baryon asymmetry
\be
\frac{n_B}{n_\gamma} = \frac{\kappa_{CP}}{10} \times 
\left(4.4 \pm 1.9 \right) \times 10^{-6},
\ee
which for the physical value $\kappa_{CP}=9.78$ is four orders of magnitude
larger than the observed asymmetry.

\section{Discussion and conclusions\label{sec_con}}

We presented first results of the simulation of cold electroweak
baryogenesis utilizing the operator (\ref{eq:CPterm0}) that arises in
the effective action of the SM from the CP violation in the CKM
matrix. In order to make this operator applicable for small Higgs
fields in the simulation, we introduced a cutoff. We chose a cutoff of
$c\Lambda^2 = 50^2$ GeV$^2$ and found a result that is four orders of
magnitude larger than the observed baryon asymmetry. If the result
would be dominated by the infrared modes of the Higgs field in the
operator one would expect according to eqs.~(\ref{eq:Wtogauge}) and
(\ref{eq:cutoff}) a scaling as $\Lambda^{-12}$. Preliminary results of
extensive computer simulations confirm that the result does rather
scale as $\Lambda^{-4}$ which we find very encouraging. A more
detailed study of the dependence of the resulting asymmetry on the
cutoff and the applicability of the operator~(\ref{eq:CPterm0}) in
general will be topic of a subsequent work~\cite{Tranberg:inprep}.

We would like to comment on why Standard Model CP violation is
operative in the proposed mechanism. Main requirement is that the
quark masses enter in a non-polynomial way in order to avoid the
Jarlskog determinant (\ref{eq_def_Jinv}) as an upper bound on CP
violation. In principle, this is easily achieved as can be seen in the
Kaon system. However, baryogenesis typically takes place in a hot
plasma and the temperature of the plasma provides a new energy
scale. The temperature effects of the plasma render most processes in
the infrared finite what makes it hard to avoid (\ref{eq_def_Jinv}) or
a similar bound depending on
temperature~\cite{Farrar:1993sp,Farrar:1993hn}. Cold electroweak
baryogenesis operates at zero temperature.

In conclusion, we find it remarkable that a cosmologically viable
baryon asymmetry could be created at a tachyonic electroweak
transition, using only Standard Model CP violation. Although technical
issues relating to the realization of low-scale inflation~
\cite{vanTent:2004rc} and the gradient expansion in this context
persist~\cite{Hernandez:2008db}, cold electroweak baryogenesis should
be considered a serious candidate scenario for explaining the baryon
asymmetry and deserves further investigation.

%%%%%%%%%%%%%%%%%%%%%%%%%%%%%%%%%%%%%%%%%%%%%%%%%%%%%%%%%%%
\section*{Acknowledgments}
%%%%%%%%%%%%%%%%%%%%%%%%%%%%%%%%%%%%%%%%%%%%%%%%%%%%%%%%%%%

We thank Jan M. Pawlowski for useful discussions. T.K. is supported by the EU FP6 Marie Curie Research \& Training
Network 'UniverseNet' (MRTN-CT-2006-035863).  A.H. is supported by
CONACYT/DAAD, Contract No.~A/05/12566. A.T. is supported by Academy of Finland Grant 114371.

%%%%%%%%%%%%%%%%%%%%%%%%%%%%%%%%%%%%%%%%%%

\end{document}